\documentstyle[graphicx,aps]{revtex}
\begin{document}

\twocolumn[\hsize\textwidth\columnwidth\hsize
\csname@twocolumnfalse\endcsname  

\title{Cold Dark Matter from Dark Energy}
\author{Aharon Davidson, David Karasik, and Yoav Lederer}
\address{Physics Department,
Ben-Gurion University of the Negev,
Beer-Sheva 84105, Israel\\
\textsf{(davidson@bgumail.bgu.ac.il)}}

\maketitle

\begin{abstract}
    Dark energy/matter unification is first demonstrated within
    the framework of a simplified model.
    Geodetic evolution of a $\Lambda$-dominated bubble Universe,
    free of genuine matter, is translated into a specific FRW
    cosmology whose effectively induced dark component highly
    resembles the cold dark matter ansatz.
    The realistic extension constitutes a dark soliton which
    bridges past (radiation and/or matter dominated) and future
    ($\Lambda$-dominated) Einstein regimes;
    its experimental signature is a moderate redshift dependent
    cold dark matter deficiency function.
\end{abstract}
\pacs{PACS numbers: }
]

Dark matter/energy is the simplest most popular ansatz invoked
by physicists, equipped with Einstein equations, when attempting
to formulate the apparent clash between contemporary theoretical
cosmology and the piling large scale observations.
The freedom which characterizes the energy/momentum section
of Einstein equations has opened the door for a vast army of
non-conventional dark particles (MACHOs are dead, long live
the WIMPs) and/or dark equations of state, none of which 
stands on solid theoretical/experimental grounds.
Even the inflationary model, which successfully tackles some
basic cosmological riddles and predicts the acoustic peaks in
the CMB power spectrum, has not shed too much light on the dark
corners.
It elegantly explains why the curvature is almost negligible,
but leaves us quite ignorant regarding the decomposition of the
almost critical total energy density.
The cosmological dark puzzle itself trifurcates into
(i) The dark energy puzzle:
What is the origin of the tiny positive cosmological vacuum
energy $\Lambda$, and is it really a constant?
(ii) The dark matter puzzle:
What is the particle content of the Universe, and why only a
small fraction of which can be detected directly? and
(iii) The dark coincidence puzzle:
Is it a mere coincidence that dark energy and dark matter
contributions are presently of the same order of magnitude?

\begin{figure}[tbp]
    \begin{center}
	\includegraphics[scale=0.4]{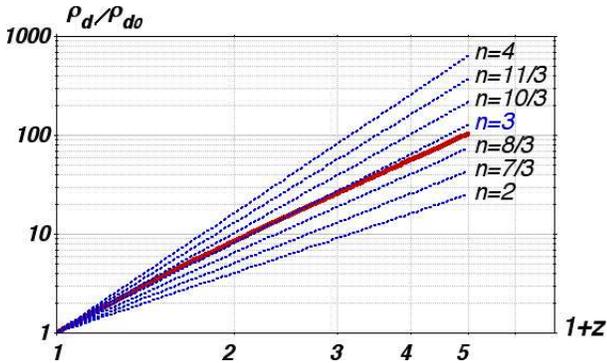}
    \end{center}
    \caption{Unified dark 'matter' (solid line) can be nicely
    approximated by the standard $n=3$ dark matter ansatz.}
    \label{rhod}
\end{figure}

\begin{figure}[tbp]
    \begin{center}
	\includegraphics[scale=0.4]{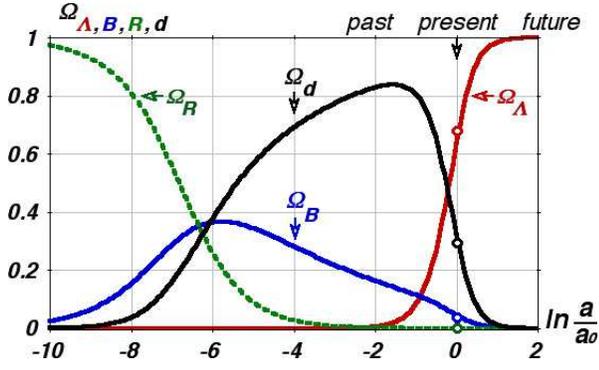}
    \end{center}
    \caption{Dark soliton connecting past ($R$-dominated)
    and future ($\Lambda$-dominated) Einstein regimes. The
    dark era extends from early $z\simeq 600$ to very recent
    $z\simeq 0.3$, peaking at $z\simeq 4.2$}
    \label{Omega}
\end{figure}

An alternative approach, that is to view the dark stuff as a
geometrical artifact owing its existence to an alternative or a
more fundamental gravitational theory, is not less speculative.
On consistency grounds, if such a theory does exist, it must
admit both a built-in Einstein limit as well as automatic
energy/momentum conservation.
Various brane theories\cite{RS,Vilenkin}, notably geodetic          
brane gravity\cite{Davidson}, fall into such a category.            
Some of them advantageously exhibit cosmological field
equations translatable into the FRW language.
In this paper, having in mind that a tenable dark companion
effectively enters the game when deviating from the Einstein
limit, we examine the idea that the apparently independent
dark puzzles share a common origin.
We refer to such an idea as dark unification.
On pedagogical grounds, before diving into the realistic
scheme, we first consider a \textit{simplified} 'dark matter
from dark energy' model of sufficient reality.
To be more specific, we confront the \textit{standard} matter
infested Universe ($\Lambda CDM$), nicely fitted by                
\begin{mathletters}
    \begin{eqnarray}
    & \rho_{standard} =
    \Lambda + \rho_{matter} ~, &\\
    & \rho_{matter} \sim a^{-3}(t) ~, & 
    \end{eqnarray}
\end{mathletters}
with a \textit{unified} bubble Universe, genuine matter free,
whose evolution, when translated into the FRW language, is
effectively governed by\cite{Davidson}
\begin{mathletters}
    \begin{eqnarray}
    & \rho_{unified} = \Lambda + \rho_{dark} ~, & \\
    & \rho^{2}_{dark}(\Lambda +\rho_{dark}) \sim a^{-8}(t) ~. &
\end{eqnarray}
\end{mathletters}
Dark unification is manifest at this level by the fact that
$\Lambda$ is the \textit{only} input parameter, with
$\rho_{dark}(\Lambda,a)$ being analytically derived.
This dark component resembles the standard dark matter ansatz
so closely, as demonstrated in Fig.\ref{rhod}, that one may 
find it hard to practically distinguish between the two models.
The corresponding Hubble plots, for instance, agree with each
other at the $1\%$ level for $z\leq 10$.
The realistic extension of the simplified model, summarized
in Fig.\ref{Omega}, is the highlight of this paper.
It is characterized by a dark soliton connecting past (radiation
and/or matter dominated) and future ($\Lambda$-dominated)
Einstein regimes, and offers an exclusive experimental signature.

Consider an embedded\cite{embedding} four dimensional brane,        
parameterized by $x^{\mu}$, floating in some given (that is
non-dynamical) $N$-dimensional background spanned by $y^{A}\,
(A=0,\ldots,N-1)$.                                                                 %
Let the brane dynamics be described by the conventional
Einstein-Hilbert Lagrangian on the brane, but
non-conventionally\cite{RT}, elevate the embedding vector           
$y^{A}(x^{\mu})$ to the level of the canonical gravitational
field.
This way, the brane metric tensor $g_{\mu\nu}(x)=g_{AB}(y)
y^{A}_{,\mu}y^{B}_{,\nu}$ becomes an induced quantity.
The field equations combined with the so-called fundamental
embedding identity then guarantee automatic energy/momentum
conservation, leaving us with\cite{Davidson}
\begin{equation}
   \left(R^{\mu\nu}-\textstyle{\frac{1}{2}}R g^{\mu\nu}
   -T^{\mu\nu}\right)\left(y^{A}_{;\mu\nu}+\Gamma^{\,A}_{BC}
   y^{B}_{,\mu}y^{C}_{,\nu}\right)=0 ~,
   \label{RTeqs}
\end{equation}
which, when using extrinsic curvatures, takes the
geometrically oriented form\cite{Carter}                            
$\left(R^{\mu\nu}-\frac{1}{2}R g^{\mu\nu}-T^{\mu\nu}\right)
K^{i}_{\mu\nu}=0$.
These $N-4$ equations describe a generalized \textit{geodetic}
motion of a bubble Universe.
Clearly, every solution of Einstein equations is necessarily a
solution of the geodetic brane equations.

Geodetic brane cosmology\cite{RTcos}, formulated by virtue of       
$5$-dim local isometric embedding in flat $M_{5}$ (or in an 
$AdS_{5}$ background, to be discussed  soon), gives rise to
a single independent equation of motion.
Upon integration, using the energy/momentum conservation law
$\dot{\rho}+3\frac{\dot{a}}{a}(\rho+P)=0$ as the integrability
condition, we get
\begin{equation}
    \rho a^{3}(\dot{a}^{2}+k)^{1/2}-
    3a(\dot{a}^{2}+k)^{3/2}=
    \textstyle{\frac{1}{\sqrt{3}}}\omega ~.
\end{equation}
The constant of integration $\omega$, identified as the
conserved bulk energy conjugate to the cyclic embedding
time coordinate $y^{0}(t)$, serves to parameterize the
deviation from the Einstein limit\cite{RTcos}.
A physicist equipped with the traditional Einstein formalism,
presumably unaware of the underlying brane physics, would
naturally re-organize the latter equation into
\begin{equation}
    \dot{a}^{2}+k =
    \textstyle{\frac{1}{3}}\left(\rho+\rho_{d}\right)a^{2} ~,
\end{equation}
squeezing all 'anomalous' pieces into dark $\rho_{d}$.
Our physicist may rightly conclude that the Einstein evolution
of the Universe is governed by a total
$\rho_{T}\equiv\rho+\rho_{d}$, rather than by plain $\rho$.
It is remarkable that the implicit formula for the dark
component\cite{Davidson}, namely                                 
\begin{equation}
    \rho_{d}^{2}\left(\rho+\rho_{d}\right)=
    \frac{\omega^{2}}{a^{8}} ~,
    \label{rho_dark}
\end{equation}
guarantees the definite positivity of the total energy density
(which cannot vanish off the Einstein limit).
We remark in passing that, in an $AdS_{5}$ background, the
above master equation is elegantly generalized into
$\rho_{d}^{2}\left(-\frac{1}{2}\Lambda_{5}+\rho+\rho_{d}\right)=
\omega^{2}a^{-8}$.
This way, if $|\Lambda_{5}|\gg\rho_{T}$, the characteristic
Randall-Sundrum
$\rho_{d}^{RS}\sim a^{-4}$ piece\cite{RScos} emerges.                
Again, as far as this paper is concerned, $\Lambda_{5}=0$.
              
The time is ripe now for the practical question:
Can a simple input $\rho$, after taking into account the
dark effect and setting $k=0$, reasonably describe the
observed Universe?
A clue may come from the 'empty' case
\begin{equation}
    \rho=0 ~~\Longrightarrow~~ 
    \rho_{T}=\rho_{d}(0,a)=\frac{\omega^{2/3}}{a^{8/3}}~,
    \label{empty}
\end{equation}
telling us that an empty bubble Universe evolves intriguingly
similar to a matter dominated FRW Universe.
However, a premature model, based on $\rho=0$ or even on
$\rho\sim a^{-3}$, falls short on realistic grounds; it would
never generate enough negative pressure
($P_{T}\geq -\frac{1}{9}\rho_{T}$, whereas
$P_{T}\leq -\frac{1}{3}\rho_{T}$ is needed) to support an
accelerating Universe.
In other words, we cannot get dark 'energy' from dark matter.
But can we still get dark 'matter' from dark energy?
To find out, let us first examine the simplest case
\begin{equation}
    \rho=\Lambda ~~\Longrightarrow~~ 
    \rho_{T}=\Lambda+\rho_{d}(\Lambda,a) ~,
\end{equation}
recognized as the geodetic brane distortion of the deSitter
model, and focus on its single positive $\rho_{d}\geq 0$ branch.
To present our results, it is crucial to fix a current
value for $\Omega_{d}\equiv\rho_{d}/\rho_{crit}$.
For the sake of definiteness, it is hereby taken to
be\cite{data} around $\Omega_{d0}\simeq 0.3$.

\begin{figure}[tbp]
    \begin{center}
	\includegraphics[scale=0.4]{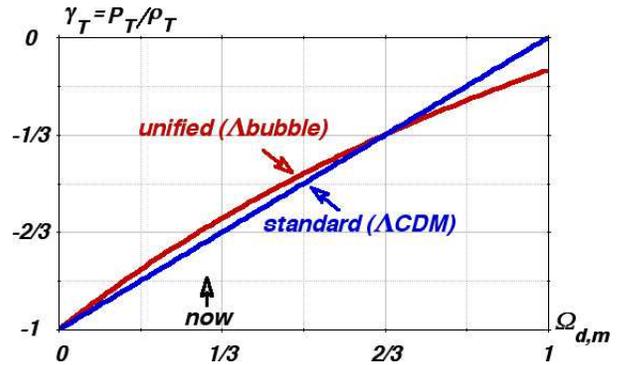}
    \end{center}
    \caption{Unified vs. Standard total equations of state.
    The standard plot can be viewed as a linear approximation
    to the curved unified plot.}
    \label{EqState}
\end{figure}

The first item on the list is the deceleration parameter 
$q=-\ddot{a}a/\dot{a}^{2}$, comfortably given by
\begin{equation}
    q=\frac{3\Omega_{d}-2}{2+\Omega_{d}} ~.
\end{equation}
For comparison, $\Lambda CDM$ comes with
$q=\frac{1}{2}(3\Omega_{m}-2)$.
Evolving from $\frac{1}{3}$ towards $-1$, unified $q$ passes
through $0$ at $\Omega_{d}=\frac{2}{3}$, and is currently
located around $q_{0}\simeq -0.5$.
To extract the dark equation of state, first derive
eq.(\ref{rho_dark}) with respect to the cosmic time to obtain
\begin{equation}
    \frac{\dot{\rho_{d}}}{\rho_{d}}+8\frac{\dot{a}}{a}
   \left(\frac{\rho_{d}+\Lambda}
   {3\rho_{d}+2\Lambda}\right) = 0 ~,
   \label{rhodot}
\end{equation}
which can now be interpreted as the dark conservation law
$\dot{\rho_{d}}+3\frac{\dot{a}}{a}(P_{d}+\rho_{d})=0$.
This, in turn, fixes the dark equation of state, which takes
the compact form
\begin{equation}
    \gamma_{d} \equiv \frac{P_{d}}{\rho_{d}}=-\frac{q}{3} ~.
    \label{darkEq}
\end{equation}
The total equation of state can be easily derived using a
similar technique, and one finds
\begin{equation}
    \gamma_{T} \equiv \frac{P_{T}}{\rho_{T}}=
    -\frac{6-5\Omega_{d}}{3(2+\Omega_{d})} ~.
\end{equation}
The comparison with the standard equation of state
$P_{T}=-(1-\Omega_{m})\rho_{T}$ is done in Fig.\ref{EqState}.
In a sense, the standard plot can be viewed as a linear
approximation to the unified plot.
A physicist ignorant of dark unification may naively
interpret $\{\Lambda+\rho_{d},\Omega_{d0}\}$ as\cite{eqState}     
$\{\rho_{X}+\rho_{m},\Omega_{m0}\}$.
In which case, the relation $(1-\Omega_{m})\gamma_{X}=
\gamma_{T}$ opens the door for $\gamma_{X0}$ to take values
below or even above $-1$.
It follows from eq.(\ref{rhodot}) that the power behavior of
the dark component can be nicely approximated by
\begin{equation}
    \rho_{d}\approx \frac{1}{a^{n}} ~,~~
    n=\frac{8}{\Omega_{d}+2} ~,
\end{equation}
for a slow varying $n$.
Tracing the evolution of the bubble Universe, the power
$n$ recovers from its past asymptotic value
$n\rightarrow\frac{8}{3}$, just below $3$, climbing
monotonically towards its future asymptotic value
$n\rightarrow 4$, currently passing $n_{0}\approx 3.5$, just
above $3$; an exact $n=3$ matter-like behavior is locally
detected at $\Omega_{d}=\frac{2}{3}$.
In other words, such a power dependence averagely resembles
the standard $n=3$ dark matter ansatz.
To sharpen this $\rho_{d}\leftrightarrow\rho_{m}$ similarity
and appreciate the integrated effect, we now leave the
pedagogical slow varying $n$ approximation, and would like
to compare the dark density ratio $\rho_{d}(z)/\rho_{d0}$
with various powers  of $(1+z)$.
This is carried out in Fig.\ref{rhod}, and comes with a clear
message.
It is amazing how fantastically close is the predicted dark
'matter' which accompanies our unified model to the standard
dark matter ansatz.
Note that whereas a small $\Lambda_{5}$ is tolerable (or even
welcome), a large $\Lambda_{5}$, due to the enhanced RS dark
piece $\sim a^{-4}$, would spoil the matter-like behavior of
$\rho_{d}$ at small $z$.
                                                            
The Hubble plot is perhaps the best tool for directly testing
the $\rho_{d}\leftrightarrow\rho_{m}$ similarity.
Can it really tell the standard cold dark matter from the
unified effectively induced dark imposter?
To find out, we have calculated the luminosity distance
\begin{equation}
    \begin{array}{l}
        \displaystyle{d_{L}(z)=
	\frac{a_{0}^{2}}{a(z)}\int_{a(z)}^{a_{0}}
        \frac{\sqrt{3}da}
	{a^{2}\sqrt{\Lambda+\rho_{d}(\Lambda,a)}}=}
	\vspace{8pt} \\
        \displaystyle{\frac{(1+z)\xi_{0}^{3/8}}{3H_{0}
	(\xi_{0}-1)^{1/4}}
        \left[\frac{8x^{5/8}F(x)}{5(x-1)^{3/4}}
        -\frac{(x-1)^{1/4}}{x^{3/8}}
        \right]_{\xi_{0}}^{\xi(z)}}~.
    \end{array}  
\end{equation}
Here, $F$ stands for the Gauss hypergeometric function
$F(x)\equiv~_{2}F_{1}(\frac{3}{4},1,\frac{13}{8},\frac{x}{x-1})$,
and $\xi(z)\geq 1$ is nothing but a root of the cubic equation
\begin{equation}
    \xi(\xi-1)^{2}=\frac{\Omega_{d0}^{2}(1+z)^{8}}
    {(1-\Omega_{d0})^{3}} ~,
\end{equation}
such that $\xi_{0}=(1-\Omega_{d0})^{-1}$.
The relevant parameter, to measure how close (numerically) are
the unified and the standard Hubble plots, is obviously the
relative luminosity distance
$(d_{L}^{unified}-d_{L}^{standard})/
(d_{L}^{unified}+d_{L}^{standard})$.
One may immediately verify, by plotting this quantity for
$\Omega_{d0}=\Omega_{m0}\simeq 0.3$ (we skip the plot due to
length limitation), that the two Hubble plots agree with each
other at the $1\%$-level for $z\leq 10$ (above and beyond the
supernova data\cite{SN}).                                            
Another rewarding exercise is to calculate the age of the
geodetically evolving $\Lambda$-bubble. 
The corresponding formula being
\begin{equation}
    \tau=\frac{1}{4}\sqrt{\frac{3}{\Lambda}}
    \left(\sqrt{\Omega_{\Lambda}}
    -\ln\frac{1-\sqrt{\Omega_{\Lambda}}}
    {1+\sqrt{\Omega_{\Lambda}}}\right) ~.
\end{equation}
Plotting the unified and the standard ages in units of constant
$\sqrt{3/\Lambda}$ rather than in the conventional $1/H$
units, makes it easier to appreciate their numerical similarity
for all $\Omega_{d,m}$. 
For the sake of clarity, however, translating to conventional
units and setting $\Omega_{\Lambda 0}\simeq 0.7$, we obtain
\begin{equation}
    \tau_{unified} \simeq \frac{0.97}{H_{0}} ~~,~~
    \tau_{standard} \simeq \frac{0.96}{H_{0}} ~,
\end{equation}
in a remarkable ${\cal O}(1\%)$ agreement, and fully consistent
with current data.

\begin{figure}[tbp]
    \begin{center}
	\includegraphics[scale=0.4]{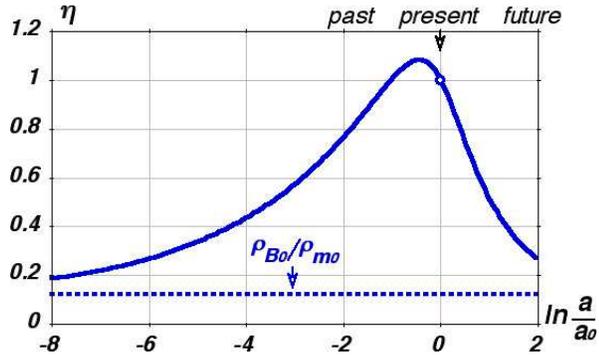}
    \end{center}
    \caption{The matter deficiency function measures the
    relative amount of 'matter' in comparison with standard
    $\Lambda CDM$.}
    \label{deficiency}
\end{figure}

The elegance of the geodetically evolving $\Lambda$-dominated
bubble Universe need not fool us.
In many respects, although being good news for the forthcoming
realistic model, which must exhibit exactly such a behavior at
the small-$z$ region where baryonic matter and radiation are
negligible, it cannot constitute the full picture.
Adding the missing ingredients to the game, that is starting
from $\rho=\Lambda+Ba^{-3}+Ra^{-4}$, the pretentiously realistic
dark component is then the positive root of
\begin{equation}
    \rho_{d}^{2}\left(\Lambda+\frac{B}{a^{3}}+\frac{R}{a^{4}}+
    \rho_{d}\right) = \frac{\omega^{2}}{a^{8}} ~.
\end{equation}
The Einstein limit is approached as
$\displaystyle
{E(\rho,a)\equiv\frac{\omega^{2/3}}{\rho a^{8/3}}\rightarrow 0}$.
Near the Einstein limit
\begin{equation}
    \rho_{T}=\rho\left(
    1+E^{3/2}+\ldots
    \right) ~.
\end{equation}
The fact that associated with the two pieces which constitute
$\rho$, namely $\Lambda$ and $Ba^{-3}+Ra^{-4}$, are powers of
$a(t)$ above and below $-\frac{8}{3}$, respectively, is crucial
for our analysis.
For $a\rightarrow\infty$, we have $\rho\simeq\Lambda a^{0}$
such that $E\sim a^{-8/3}\rightarrow 0$, signaling the future
Einstein limit.
For $a\rightarrow 0$, on the other hand, we have $\rho\simeq
Ba^{-4}$ leading to $E\sim a^{4/3}\rightarrow 0$, marking the
past Einstein limit ($E\sim a^{1/3}\rightarrow 0$ for $R=0$),
with $\rho_{d}$ serendipitously mimicking a curvature term.
The complete cosmological evolution, subject to the present 
data\cite{data,SN}, is depicted in Fig.\ref{Omega}.
It describes a bubble Universe in geodetic transition from
radiation to $\Lambda$ domination, and is characterized by a dark
soliton which connects these two distinct (past and future)
Einstein regimes.
The dark effect peaks at
\begin{equation}
    1+z_{max}=\left(\frac{8\Omega_{\Lambda 0}}
    {\Omega_{B0}}\right)^{1/3} ~.
\end{equation}
Special attention should be devoted to the combination 
$\rho_{B}+\rho_{d}$, the unified analog of the standard
$\rho_{m}$.
At any given $a(t)$, the ratio
$\eta\equiv(\rho_{B}+\rho_{d})/\rho_{m}$ measures the relative
amount of 'matter' in comparison with standard wisdom.
In other words, the $\Lambda CDM$-like FRW evolution is
governed by $\Lambda+\eta(a)a^{-3}$.
Normalized to unity today, the moderate $\eta$-function (plotted
in Fig.\ref{deficiency}) tends to $\rho_{B0}/\rho_{m0}$ at very 
early (and very late) times.
Such a detailed deficiency function of matter density appears
to be the main testable prediction of dark unification.

To summarize, the name of our game is simplicity.
Our idea is to trade the standard cold dark matter ansatz by a
fundamental underlying physical principle.
In this respect, simplicity is spelled out as dark unification.
At the simplified level, the sole input of our geodetic gravity
model is the cosmological constant $\Lambda$, and nothing but
$\Lambda$.
The corresponding $\Lambda$-bubble Universe model is then the
exact analog of deSitter model of Einstein gravity; both models
are associated with the one and the same gravitational action,
but conceptually differ from each other by the choice of the
canonical gravitational fields.
At the realistic level, some genuine baryonic matter and
radiation are added to the game.
Artifact dark 'matter' makes its effective entrance when
attempting to formulate bubble Universe evolution by means
of the traditional Einstein equations.
In particular,
(i) The amount of such dark 'matter' is controlled by the
conserved bulk energy which parametrizes the deviation from
the Einstein limit,
(ii) The functional behavior of the emerging dark 'matter'
soliton is fully dictated by the theory,
(iii) The numerics involved highly resemble the standard cold
dark matter ansatz, and
(iv) The dark era extends from $z\simeq 600$ until today.
Dark unification does leave, however, an exclusive experimental
fingerprint in the shape of a moderate $z$-dependent matter
deficiency function; its consequences are currently under
investigation.

Clearly, reflecting its underlying first principles, the
universality of dark unification must be absolute.
Thus, our major challenge is a galactic scale realization
of the cosmological dark 'matter' idea, presumably but not
necessarily in the form of some dark soliton bridging two
Einstein regimes.
The main theoretical obstacle at the moment is the exact
radially symmetric (and time dependent) geodetic brane analog
of the Schwarzschild or Schwarzschild-deSitter solution, which
is still unknown.
An important progress\cite{Newton} in this direction, namely
the recovery of the Newtonian limit in an 'empty' (in the sense
of eq.\ref{empty}) dark cosmological background, can already
be reported. 

Thanks to D. Eichler, E. Guendelman, and R. Brustein for their
constructive comments, and to I. Maor for conducting a Hubble
plot likelihood analysis.

\end{document}